\documentclass[a4paper,9pt]{article}
\usepackage{fullpage}
\usepackage{mathpazo}
\usepackage[frozencache]{minted}
\usepackage{graphicx}
\usepackage{amsmath}
\usepackage{caption}
\usepackage{subcaption}
\usepackage[backend=biber]{biblatex}
\usepackage{url}
\usepackage[]{hyperref}

\begin{document}

\title{Interprocess Communication in FreeBSD 11: \\ Performance Analysis}
\author{A.~H.~Bell-Thomas \\ \small{Alexander.Bell-Thomas@cl.cam.ac.uk}}
\date{}

\maketitle

\thispagestyle{empty}

\begin{abstract}
Interprocess communication, IPC, is one of the most fundamental functions of a modern operating system, playing an essential role in the fabric of contemporary applications. This report conducts an investigation in FreeBSD of the real world performance considerations behind two of the most common IPC mechanisms; \textit{pipes} and \textit{sockets}. A simple benchmark provides a fair sense of effective bandwidth for each, and analysis using \textit{DTrace}, hardware performance counters and the operating system's source code is presented. We note that \textit{pipes} outperform \textit{sockets} by 63\% on average across all configurations, and further that the size of userspace transmission buffers has a profound effect on performance --- larger buffers are beneficial up to a point ($\sim$ 32-64 KiB) after which performance collapses as a result of devastating cache exhaustion. A deep scrutiny of the probe effects at play is also presented, justifying the validity of conclusions drawn from these experiments.

% \bigskip

\end{abstract}

\clearpage

\setcounter{page}{1}

\section{Introduction}
\paragraph{} One of the most fundamental mechanisms of any modern operating system is interprocess communication (IPC). This is especially true in environments based on UNIX, where it is standard for a large number of disparate applications to be stitched together to fulfil a task. FreeBSD provides a wide range of IPC implementations, such as shared lock management and memory, but two mechanisms in particular dominate this space; pipes and sockets. This report shall investigate and compare both.

\paragraph{} Both pipes and sockets are inherited from the POSIX\footnote{Portable Operating System Interface.} standard, and although they are internally very different from one another they are both presented to users via generic file-descriptors. Pipes facilitate the sending of unidirectional, unnamed byte-streams between entities; this is a very popular dataflow system when using command-line tools. They require a common parent process to set up the communication channel. Notably, before 4.2BSD pipes were implemented via the filesystem and in later versions using sockets. FreeBSD no longer uses socket-backed pipes, instead opting for a separate optimised implementation directly on top of the virtual memory system.~\cite{10.5555/2659919} Sockets on the other hand are designed for bidirectional communication and are highly adaptable, affording support for complex data structures such as packets when interfacing with a network. This report shows that pipes outperform sockets due to implementation optimisations in how they manage memory, and additionally that larger buffer sizes will improve performance up to a point, after which it degrades.

\paragraph{} The focus of this report is to determine the performance characteristics of these two IPC mechanisms when used across both two threads and two processes. The analysis conducted considers a number of influential factors; particular focus is given to the internal structure and interactions of the kernel's components, as well as system behaviours at both architectural and micro-architectural levels. Section 2 details the experimental setup and methodology used, including a set of hypotheses that the analysis (Section 3) attempts to resolve. Section 4 present the conclusion to this investigation.

\section{Experimental setup and methodology}
\paragraph{} The research questions this report aims to answer are as follows.

\begin{enumerate}
    \item How does increasing IPC buffer size uniformly change performance across IPC models?
    \item How does changing the IPC buffer size affect the architectural and micro-architectural aspects of cache and memory behaviour?
    \item Can we reach causal conclusions about the scalability of the kernel’s pipes and local socket implementations given evidence from processor performance counters?
\end{enumerate}

\paragraph{} These questions were used to derive the guiding set of hypotheses from which the experiments presented in this reports were derived. At all points the probe effect is considered. The benchmark was tested unaltered, but only one operation mode, \texttt{2thread}, with a single static total IPC size, 16 MiB, was of interest.

\subsection{Hypotheses}

\paragraph{\textit{Hypothesis 1}: Increasing the buffer size available to each IPC mechanism will improve their performance up to a maximum, after which it will degrade.} Regardless of the IPC mechanism used, data has to be transferred block by block between sending and receiving entities; this is facilitated via buffers at the both the receiver and sender. In the case of sockets there is an additional in-kernel intermediary buffer, the size of which can be manipulated by applications (the \texttt{-s} flag in the benchmark). The buffer size chosen should be an essential factor of real-world performance; the \textit{best} choice will most likely depend on the IPC mechanism and other system conditions. Logically, one would expect a larger buffer size to better amortise the effect of \textit{syscall} overhead for large transmissions, but as shown in the first lab report this must be balanced with adverse effects on the processor's cache memories.

\paragraph{\textit{Hypothesis 2}: Pipes will yield better performance than sockets for local communication due to their specific VM-optimisations.} Sockets are notably far more versatile than pipes, theoretically putting them at a disadvantage in IPC performance testing. They need to interface not only across threads and processes but additionally across networks, a medium that brings a very different set of constraints and challenges. Pipes on the other hand are known to have a IPC implementation separate to that of sockets, giving room for context-aware optimisation directly on top of FreeBSD's VM system.

\paragraph{\textit{Hypothesis 3}: The use of instrumentation tools such as \textit{DTrace} and PMCs will adversely affect the benchmark's performance, but overall the results will demonstrate a consistent shape, especially considering their inflection points.}
\subsection{Background}

\paragraph{} All experiments, and therefore results recorded, were conducted on a BeagleBone Black (revision C) board.~\cite{bbb, BBBSRMpd2} It is equipped with a 32-bit ARM Cortex-A8 (AM335x, 1GHz, 32 KB L1 cache, 256 KB L2 cache),~\cite{AM335xAR80} 512 MB of DDR3L RAM (800MHz, 1.6 GiB/second of memory bandwidth), and 4GB of eMCC flash memory. The operating system used is FreeBSD 11.0.0;\footnote{Modified to run on the BeagleBone Black (\texttt{version d508cb8(release/11.0.0)-dirty}).} this comes with native \textit{DTrace} support. The persistent disk used by the system is a 8GB microSDHC card. The root of the filesystem was mounted using read-only mode to ensure that no unexpected/accidental activity could affect the OS installation; the writable \texttt{/data} partition was used for running trials and recording the results. The machine was accessed locally, using both a USB serial connection and SSH over Ethernet.

\paragraph{} The heart of our investigation of the hypothesis set was an I/O benchmark program written by Robert N.M. Watson. The benchmark is able to measure IPC across both threads and processes via either pipes or sockets: the corresponding POSIX APIs are \texttt{pthread\_create}, \texttt{fork}, \texttt{pipe}, and \texttt{socketpair} respectively. For sockets, the kernel's internal buffer size can be changed using \texttt{setsockopt}; the benchmark exposes this via the \texttt{-s} flag, setting it to be the same size as the userspace buffers. In addition to recording the effective IPC bandwidth seen using a particular configuration it enabled CPU performance counters (PMCs) to be queried to further enlighten our understanding of the system's behaviour. To guarantee that \texttt{read} and \texttt{write} performed full, not partial, operations, the system was configured to increase its \texttt{kern.ipc.maxsockbuf} value to 32 MiB, greater than the largest buffer size tested against. 

\paragraph{} This report's analysis of the benchmark's behaviour relies primarily on \textit{DTrace} and PMCs; these were coordinated both using shell scripts and an IPython Jupyter Notebook for data collection. The probe effect of these investigation tools is detected and measured with a number of additional experiments; the purpose of this is to verify that the benchmark's behaviour was not significantly altered. This will be discussed in depth in §~\ref{sec:probe}.

\paragraph{} Unless stated otherwise, all results presented use a sample size of at least 30 datapoints. The first datapoint for each configuration run was discarded to ensure the system was in a steady-state before recording measurements. Graphs plot the median of these samples, as well as their IQR ($25^{th}$ to $75^{th}$ percentile) as a notion of error. Statistical significance tests make use of the paired Wilcoxon signed-rank test: this was chosen over the traditional Student's $t$-test because it does not assume the underlying data is drawn from a Gaussian distribution, something that cannot be guaranteed in this context.

\paragraph{} Three distinct sets of experiments were performed to investigate the three experimental hypotheses --- software-based instrumentation to analyse kernel performance, hardware-based instrumentation to assess the impact of the processor's microarchitecture, and a statistical analysis of both to investigate the impact of probe effect.

\section{Results and Discussion}

\begin{figure}
    \centering
    \includegraphics[width=\textwidth]{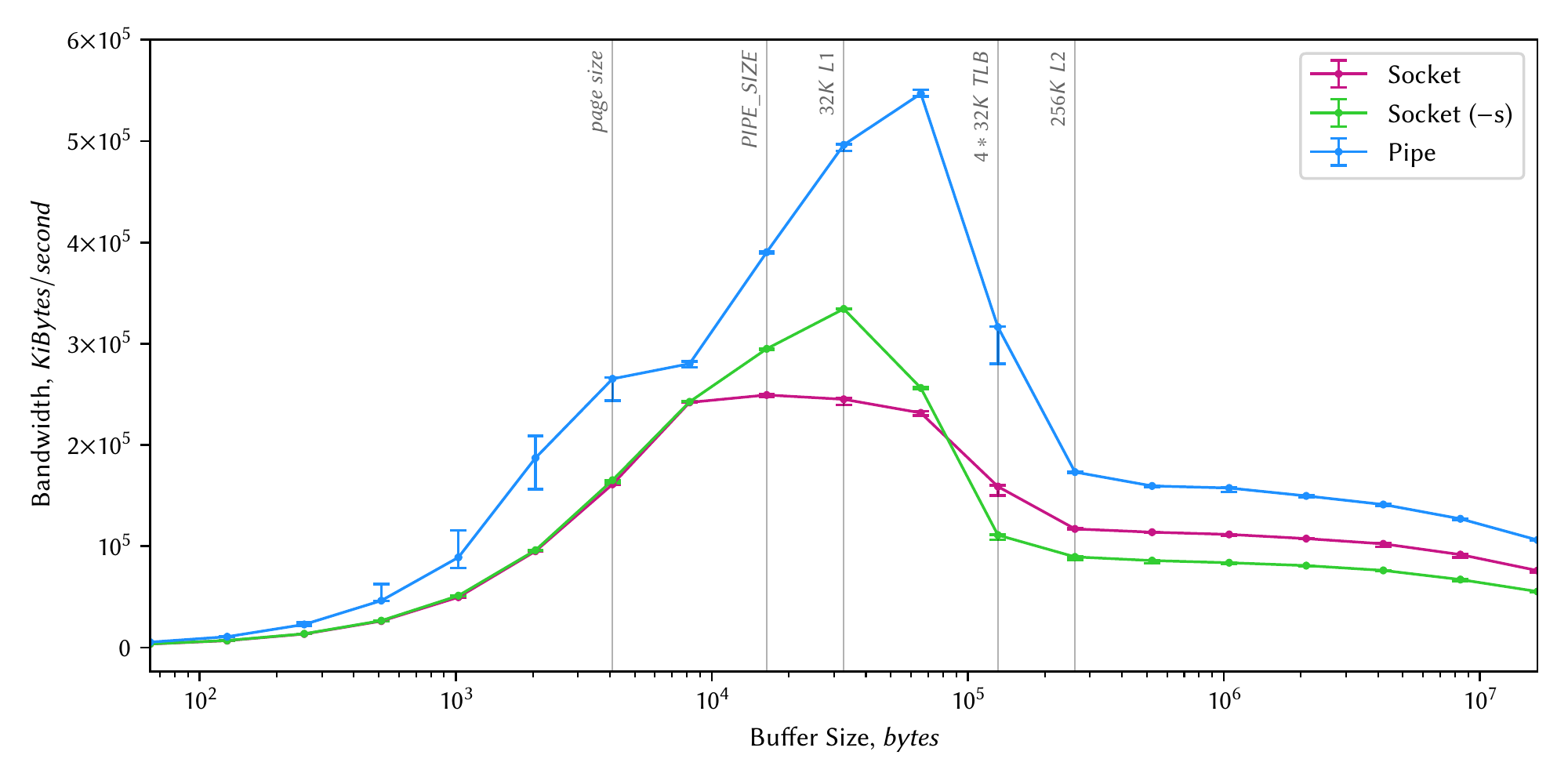}
    \caption{Effective bandwidths produced by the \texttt{ipc-static} benchmark for different IPC types and configurations as the buffer sizes used changes (no instrumentation).}
    \label{fig:uninstrumented-bandwidth}
\end{figure}

\paragraph{} In this section the results of the experiments performed will be presented, backing a discussion and evaluation of the difference between \texttt{pipe} and \texttt{socketpair} for IPC.

\subsection{Impact of the Kernel}
 
\paragraph{} Figure~\ref{fig:uninstrumented-bandwidth} presents the results produced by the benchmark across a range of buffer sizes and using each of the three IPC mechanism configurations. The \texttt{-s} qualifier denotes that the size of the kernel buffer was modified to mirror the size of the user space buffers. From first impressions there is a clear trend in the data: regardless of the IPC model, performance increases with buffer sizes up to a point ($\sim$32 to 64 KiB), after which it begins to decrease substantially. The initial increase in performance can fairly easily be attributed to a decrease in the number of \texttt{read()} syscalls given the total I/O size is fixed. This behaviour is directly in line with the high-level conceptual model presented in \textit{Hypothesis 1}. Figure~\ref{fig:uninstrumented-bandwidth} however contains a number of inflections points hinting at more subtle effects at play. To decipher this we will now consider each IPC mechanism in greater depth using \textit{DTrace} and direct source code analysis.

\subsubsection{Pipes}

\paragraph{} Pipes are a rather restricted transport mechanism, merely providing a way of sending ordered byte-streams between entities. Notably this representation is far less versatile that \texttt{socketpair}. To get a sense for how pipes behave in FreeBSD we first look at the source implementation and associated documentation. An essential observation to be gained from \texttt{kern/sys\_pipe.c}\footnote{\url{http://fxr.watson.org/fxr/source/kern/sys_pipe.c?v=FREEBSD-11-0}} is that \texttt{pipe} can operate in one of two modes; \textit{'small'} or \textit{'large'} write mode. 

\paragraph{} In \textit{'small'} write mode, when the buffer size if smaller than \texttt{PIPE\_MINDIRECT} (8 KiB),\footnote{Refer to \url{http://fxr.watson.org/fxr/source/sys/pipe.h?v=FREEBSD-11-0}.} a \textit{'normal'} buffer is provided through the kernel. For write sizes between \texttt{PIPE\_MINDIRECT} and \texttt{PIPE\_SIZE} (16 KiB) the sending process pins the underlying VM pages for the receiver to copy from directly; such optimisations are key to \texttt{pipe} demonstrating greater performance that \texttt{socketpair}, as this removes unnecessary copying via a kernel buffer.

\paragraph{} The greatest observed throughput for \texttt{pipe} occurred with a buufer size of 64 KiB, as can be clearly seen in Figure~\ref{fig:uninstrumented-bandwidth}. A contributing factor towards this is \texttt{pipe}'s resizing mechanism,\footnote{Refer to \texttt{kern/sys\_pipe.c}, lines 1079-1088.} which, if enabled,\footnote{It is on our system (\texttt{sysctl kern.ipc.piperesizeallowed $\rightarrow$ 1}).} will increase the default \texttt{PIPE\_SIZE} from 16 KiB to a maximum of \texttt{BIG\_PIPE\_SIZE} (64 KiB) in increments. This is visible via \textit{DTrace}'s \texttt{syscall::read*:entry/return} probes, which clearly showed all buffer sizes beyond 64 KiB resolving to 256 \texttt{read()} calls of size 64 KiB.

\paragraph{} Using \textit{DTrace} we further validated a number of details about the behaviour of \texttt{pipe}. Using the \texttt{fbt::pipe\_write:entry} probe descriptor the following data was extracted from \texttt{arg0} (\texttt{struct file *fp}) and \texttt{arg1} \texttt{(struct uio *uio)}).\footnote{Refer to \texttt{kern/sys\_pipe.c}, line 1039.}

\begin{quote}
    \texttt{($\alpha$) struct pipebuf pipe\_buffer \{ ...; u\_int size = 0x4000; ... \}} \\
    \texttt{($\beta$) struct uio uio \{ ...; ssize\_t uio\_resid = 0x1000000; ... \}} \\
\end{quote}
\vspace{-8mm}

\paragraph{} ($\alpha$) provides the initial size of default buffer given to the \texttt{pipe} on creation, \texttt{PIPE\_SIZE}; this verifies our conclusions from the source file. The significance of ($\beta$) is slightly more subtle. This extract was taken from a run with buffer set to 16 MiB (\texttt{=0x1000000}), and this \texttt{uio\_resid} value declares the number of bytes left to process when copying commences; this confirms that the entirety of the userspace buffer passed into \texttt{write()} is processed in one shot by \texttt{pipe\_write()}, eliminating the need for an \texttt{-s} option as we have for sockets.

\subsubsection{Sockets}
\paragraph{} The implementation of \texttt{socketpair} is far more complicated than that of \texttt{pipe}, something that is rather unsurprising given how versatile the POSIX standard forces it to be. Notes in \texttt{kern/uipc\_socket.c} and \texttt{kern/uipc\_usrreq.c} enlighten a handful of the difficulties faced, especially in the face of \textit{'ancillary data'}; credentials, file descriptors, and even, one layer deeper, other sockets themselves may potentially be passed over a socket, requiring additional considerations such as a specialised garbage collector for dead sockets. Importantly, FreeBSD's implementation does not lend itself towards interoperability between local sockets and shared memory (or other VM optimisations); this is undoubtedly key to its poorer performance when compared to \texttt{pipe} in Figure~\ref{fig:uninstrumented-bandwidth}.

\paragraph{} An important observation from Figure~\ref{fig:uninstrumented-bandwidth} is that both \texttt{socketpair} configurations achieve their maximum throughput earlier than \texttt{pipe}, also behaving differently to one another after the 8 KiB mark. This was initially investigated using the \textit{DTrace} \texttt{syscall::read*/write*:entry/return} probes. All buffer sizes larger than 8 KiB resolved to 2048 \texttt{read()} calls of 8 KiB when the \texttt{-s} flag was not used, explaining its early plateau in Figure~\ref{fig:uninstrumented-bandwidth}. This limitation was not observed when the \texttt{-s} flag was used.

\paragraph{} Deeper \textit{DTrace} instrumentation was performed to further dissect this behaviour, targeting specifically the \texttt{soreceive\_generic()} method (sockets' internal version of \texttt{read()}) in \texttt{kern/uipc\_socket.c}. Using the \texttt{fbt::soreceive\_receive:entry} probe the following data was extracted.

\begin{quote}
    \texttt{($\gamma$) struct sockbuf so\_rcv \{ ...; u\_int sb\_hiwat = 0x2000; ... \}} \\
    \texttt{($\delta$) struct sockbuf so\_snd \{ ...; u\_int sb\_hiwat = 0x2000; ... \}} \\
\end{quote}
\vspace{-5mm}

\paragraph{} Both ($\gamma$) and ($\delta$) depict the \textit{high watermark}, or maximum \texttt{char} count supported, of the \texttt{sockbuf}\footnote{\url{http://fxr.watson.org/fxr/source/sys/sockbuf.h?v=FREEBSD-11-0}} buffer structures used for both sending and receiving sockets. Given that \texttt{0x2000} = 8 KiB this is a highly plausible explanation for the plateau in \texttt{socketpair} performance between 8 KiB and 64 KiB. The \texttt{-s} flag manually sets the high and low watermarks (\texttt{int sb\_lowat}) to the size of the benchmark's buffer; this was verified using the same \textit{DTrace} probe.

% \paragraph{} Further investigation will be detailed in §~\ref{sec:hw} with the help of PMCs.

\begin{figure}
    \centering
    \begin{subfigure}[b]{0.99\textwidth}
         \centering
        \caption{Architecturally originated memory reads (\texttt{MEM\_READ})}
         \includegraphics[width=\textwidth]{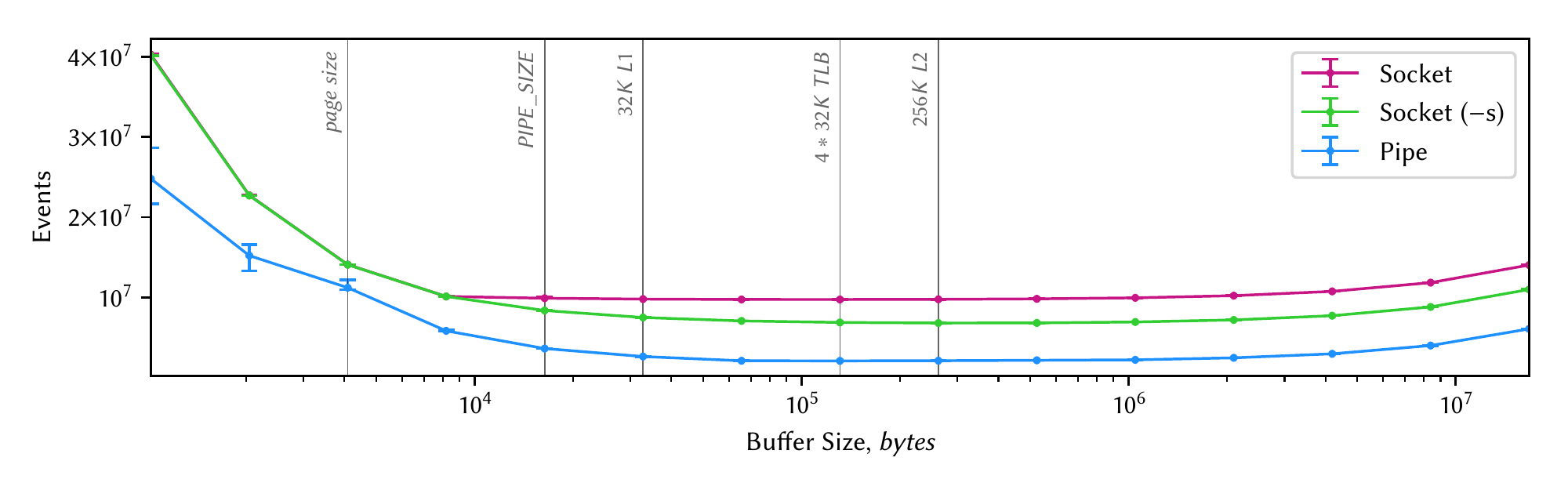}
         \label{fig:pmc-a}
    \end{subfigure}
    \newline
    \begin{subfigure}[b]{0.99\textwidth}
         \centering
        \caption{AXI bus: architectural reads (\texttt{AXI\_READ})}
         \includegraphics[width=\textwidth]{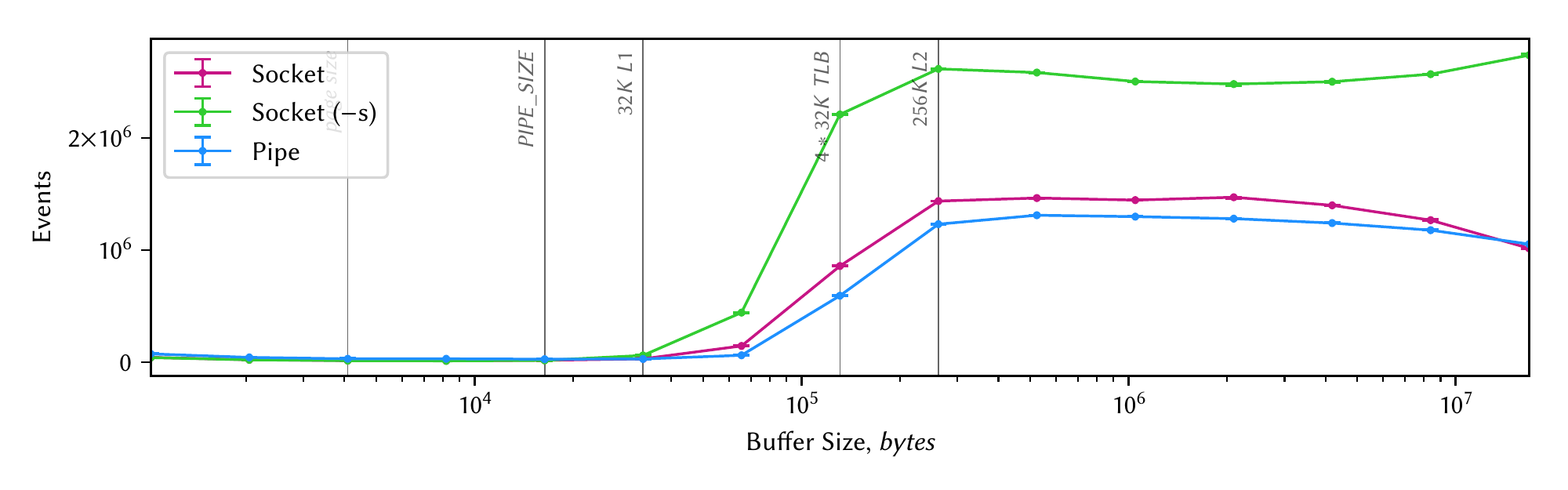}
         \label{fig:pmc-b}
    \end{subfigure}
    \newline
    \begin{subfigure}[b]{0.99\textwidth}
         \centering
        \caption{AXI bus: average clock cycles per instruction (\texttt{CLOCK\_CYCLES/INSTR\_EXECUTED})}
         \includegraphics[width=\textwidth]{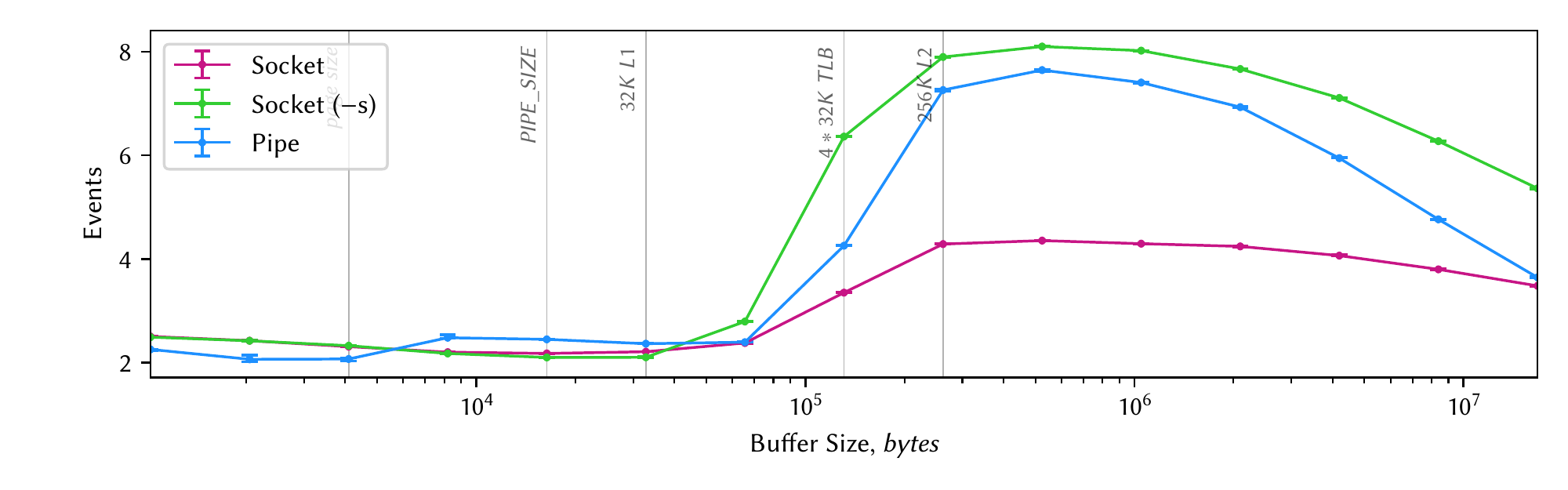}
         \label{fig:pmc-c}
    \end{subfigure}
    \newline
    \begin{subfigure}[b]{0.99\textwidth}
         \centering
        \caption{L2 cache: average clock cycles per access ($\texttt{L2\_ACCESS/CLOCK\_CYCLES}^{-1})$}
         \includegraphics[width=\textwidth]{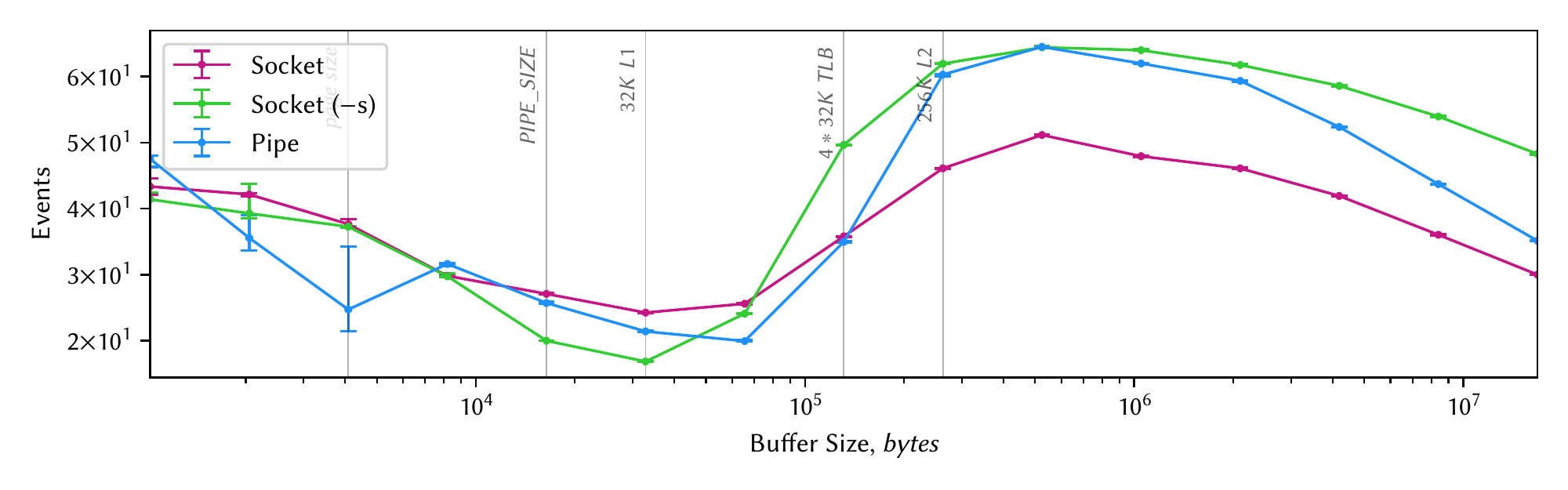}
         \label{fig:pmc-d}
    \end{subfigure}
    \vspace{-18pt}
    \caption{Selected plots of PMC attributes as buffer size increases.}
    \label{fig:pmc}
\end{figure}

\subsection{Microarchitectural Investigation}
\label{sec:hw}

\paragraph{} In the previous section we discussed various aspects of the kernel's behaviour that affect the benchmark's performance as seen in Figure~\ref{fig:uninstrumented-bandwidth}. A handful of the data's inflection points have not yet been explained, thus PMCs are now used to extend our understanding further by reasoning at the microarchitectural level. Figure~\ref{fig:pmc} presents four selected PMC attributes, demonstrating their behaviour as buffer size increases.

\paragraph{} Figure~\ref{fig:pmc-a} depicts the \texttt{mem} PMC counter's \texttt{MEM\_READ} attribute; this can be used to give an approximate indicator of the number architectural read operations. Frustratingly this cannot be directly translated into the number of bytes read. This is, we believe, due to specialised ARM instructions such as \texttt{LDRD}, which loads from two locations simultaneously in one operation --- \texttt{LDRD} is used at various points in FreeBSD's ARM implementation, including \texttt{memcpy}.\footnote{Refer to \url{contrib/cortex-strings/src/arm/memcpy.S}} However from this we are able to very clearly pick out the point at which \texttt{socketpair} with and without the \texttt{-s} flag deviates (8 KiB), reaffirming the behaviour reported by \textit{DTrace}. As the number of architectural reads directly translates as I/O load on the system, lower values are better. The VM optimisations of the \texttt{pipe} implementation found in the FreeBSD source can be seen coming into play, with a far lower impact in hardware. Additionally, other inflection points can also be seen; \texttt{socketpair} with the \texttt{-s} flag plateaus at 32-64 KiB, and \texttt{pipe} at 64 KiB, mirroring the story told by the bandwidth readings reported in Figure~\ref{fig:uninstrumented-bandwidth}. \texttt{MEM\_READ} has shown itself to be a fairly reliable proxy for bandwidth performance at lower buffer sizes, especially as it reliably exposes software behaviours, encapsulating the bare-metal requests exiting the kernel.

\paragraph{} The sudden drop in bandwidth performance after the 32 and 64 KiB marks has yet to be explained. With this in mind, the \texttt{axi} PMC tells an interesting story, bringing the root cause of the phenomenon to light. Figures~\ref{fig:pmc-b} and \ref{fig:pmc-c} show the \texttt{AXI\_READ} and \texttt{INSTR\_EXECUTED/CLOCK\_CYCLE} attributes respectively. The \texttt{axi} PMC measures activity on the chip's AXI bus, which is responsible for actually transporting data to physical I/O devices or DRAM, therefore capturing requests which are not satisfied by the CPU's cache hierarchy. Figure~\ref{fig:pmc-c} has been included as it serves as a useful metric of the average relative expense of each AXI transaction.

\paragraph{} A vital observation to make to explain the sharp drop in bandwidth performance is that as performance decreases in the 32-256 KiB buffer size range, both the number and relative expense of AXI operations increase drastically. Taking the readings observed for \texttt{pipe} as an example, and given no other PMC attributes reveal anything of particular note, it is highly likely that these observed behaviour on the AXI bus are direct indicators of the core issue causing performance to crash.

\paragraph{} There is no L3 cache in the BeagleBone Black's Cortex-A8 processor, making the L2 cache the last level of on-chip memory. Sadly the Cortex-A8 does not expose a PMC for measuring the number of L2 cache misses, but as a proxy Figure~\ref{fig:pmc-d} plots the average clock cycles per L2 cache operation; this will implicitly expose the number of misses as operations take longer whilst fetching data from DRAM. We can see that this proxy L2 miss metric aligns almost exactly with the increased strain seen on the AXI bus, giving credence to the assertion that the observed performance collapse is directly caused by L2 cache exhaustion. This further explains why both the increase in load and decrease in performance cease to change so drastically when the marked L2 limit is hit; this is the first truly degenerate case as the working set no longer fits in the last-level cache. For this reason it is clear to see how this impacts all IPC mechanisms in roughly the same way.

\subsection{Probe Effect}
\label{sec:probe}

\begin{figure}[t]
    \centering
    \begin{subfigure}[b]{0.99\textwidth}
         \centering
        \caption{Effective bandwidths achieved by the benchmark under various and no instrumentation.}
         \includegraphics[width=\textwidth]{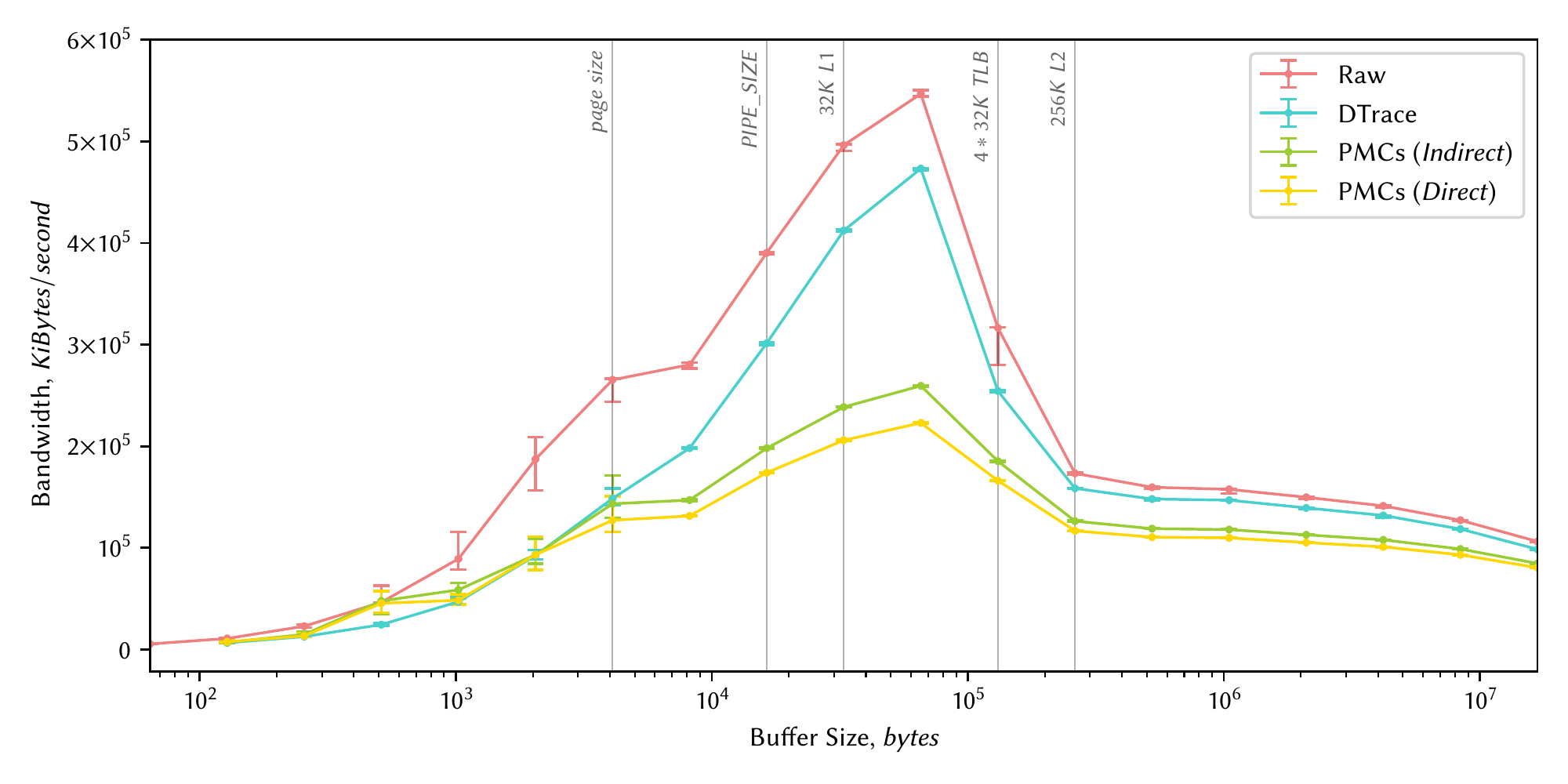}
         \label{fig:probe-a}
    \end{subfigure}
    \newline
    \begin{subfigure}[b]{0.99\textwidth}
         \captionsetup{justification=centering}
        \caption{Paired Wilcoxon signed-rank tests comparing results from instrumented runs to \\ uninstrumented performance after normalisation.}
         \includegraphics[width=\textwidth]{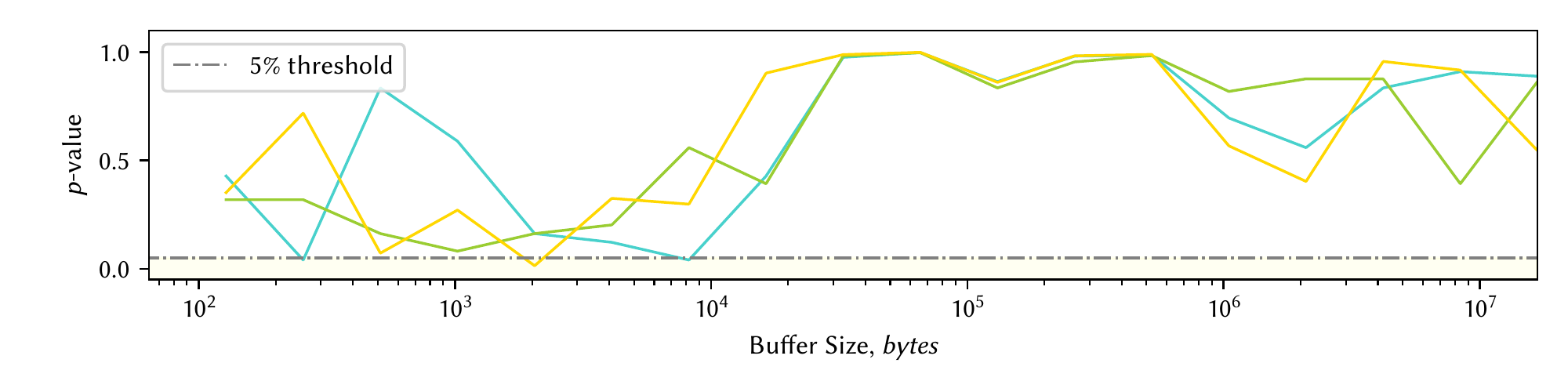}
         \label{fig:probe-b}
    \end{subfigure}
    \vspace{-15pt}
    \captionsetup{justification=centering}
    \caption{Plots examining the impact of probe effect from both \textit{DTrace} and PMCs on the \texttt{ipc-static} benchmark. The results shown are for \texttt{pipe}, although \texttt{socketpair} behaves nearly identically.}
    \label{fig:probe}
\end{figure}

\paragraph{} In all experiments discussed thus far the utmost care was taken to minimise the effect of any probe effect --- no \textit{DTrace}-instrumented runs were included in the final dataset, and PMCs were only activated when their output was required for the experiment. \textit{Hypothesis 3} stated that instrumentation was going to affect results but not the overall shape of the graph; Figure~\ref{fig:probe} plots the results of a set of experiments designed to test this. Only the results for \texttt{pipe} are plotted, but \texttt{socketpair} demonstrated very similar behaviours.

\paragraph{} The \textit{Raw} line depicts the performance of the benchmark with no instrumentation enabled. \textit{DTrace} plots performance when using the \texttt{syscall::read*/write*:entry/return} probes. Two \textit{PMC} lines are shown here; direct (\texttt{l1d, mem, axi, tlb}) and indirect (\texttt{l1i, l2}).\footnote{This information comes from the briefing document for the third L41 lab session.} This refers to whether a PMC can or cannot be directly measured on a Cortex-A8. Personally there are two surprises here;
\begin{enumerate}
    \item PMCs, despite being accelerated in hardware, perform far worse that (software-based) \textit{DTrace}. Obviously this is not a fair comparison as the two measure different things, but the difference helps to illustrate the hit required to gather such low-level runtime data.
    \item The PMCs that are not directly countable by the processor outperform those that are --- the median difference is 7.8\% (IQR: 6.3\% $\rightarrow$ 12.5\%). We speculate that this was caused by the data recorded for direct PMCs being dragged down by the \texttt{mem} and \texttt{axi} counters as they are not natively recording in the processor.

\end{enumerate}

\paragraph{} From Figure~\ref{fig:probe-a} it is clear that the first half of \textit{Hypothesis 3} is correct; instrumentation had a tremendous affect on results produced by the benchmark. A slightly more interesting question is how similar these lines are --- we hope that the basic shape of all are similar, and in an ideal world are the same line offset by constant scale factors $k_1, k_2, k_3$. To test this, each line was normalised against its mean value across all buffer sizes\footnote{This is admittedly imperfect in the general case --- here, however, the results were manually inspected before being deemed acceptable for this purpose.} --- this should conceptually bring each in-line with each other, eliminating constant scale factors. Each of the three normalised instrumentation lines were then compared against the \textit{Raw} plot using a paired Wilcoxon signed-rank test to determine whether their differences are statistically significant. These are plotted in Figure~\ref{fig:probe-b} with a 5\% significance level. The null-hypothesis (that normalised lines are not significantly different) is rejected at only three points when buffer sizes are small --- larger buffer sizes fare well, demonstrating high $p$-values.

\paragraph{} Overall, this give satisfactory credibility to \textit{Hypothesis 3} in this context.

\section{Conclusion}
\paragraph{} This report investigated the effective bandwidth of both pipes and sockets in FreeBSD 11. Drawing on the source code of the OS, its documentation, \textit{DTrace} profiling and hardware performance counters the three initial experimental hypothesis have been shown to be generally correct. Specifically, we have shown:

\begin{enumerate}
    \item Increasing buffer sizes improves performance for all IPC mechanisms up to a point ($\sim$ 32-64 KiB), after which it degrades. Thus the optimal cache size depends on a wide range of factors, but overhead amortisation is fundamentally in contention with the cache performance as buffers grow larger. \textbf{\textit{Hypothesis 1}} can therefore be soundly accepted.
    \item Pipes yield better performance than sockets for local communication due to specific memory optimisations. Further, sockets are inherently constrained by their versatile design and independent in-kernel buffer, adding up to a far less scalable IPC solution. Virtual memory optimisations for \texttt{pipe} vastly decrease the expense of performing operations, as shown in Figure~\ref{fig:pmc-a}, for example --- this indicates a far more scalable solution for local IPC than \texttt{socketpair}. \textbf{\textit{Hypothesis~2}} is also accepted confidently, citing evidence from FreeBSD's source code and observed PMC attribute results to assert that VM-optimisations were a major component of \texttt{pipe}'s speed.
    \item Instrumentation affects performance, but, to a great extent, neither the shape nor inflection points of results. \textbf{\textit{Hypothesis 3}} is therefore tentatively accepted; there is no strong evidence to reject it, although for smaller buffer sizes there may be small, but significant, artefacts.
\end{enumerate}

% \paragraph{}

\clearpage
% \bibliographystyle{unsrtnat}
% \bibliography{refs}
\printbibliography

% \clearpage
% \appendix

\end{document}